\documentclass[3p,times,procedia]{elsarticle}
\usepackage{nupha_ecrc}

\volume{00}

\firstpage{1}

\journalname{Nuclear Physics A}

\runauth{Y.~Fujimoto}

\jid{nupha}

\jnltitlelogo{Nuclear Physics A}

\usepackage{amsmath,amssymb}

\usepackage[figuresright]{rotating}

\newcommand{\dd}{\langle dd \rangle}

\begin{document}

\begin{frontmatter}



\dochead{XXVIIIth International Conference on Ultrarelativistic Nucleus-Nucleus Collisions\\ (Quark Matter 2019)}

\title{Continuity from neutron matter to color-superconducting quark matter with ${}^{3} P_{2}$ superfluidity}


\author{Yuki~Fujimoto}

\address{Department of Physics, The University of Tokyo, 7-3-1 Hongo, Bunkyo-ku, Tokyo 113-0033, Japan}

\begin{abstract}
I clarify how the concept of quark-hadron continuity, which was
previously considered in the context of the asymptotic color-flavor
locked phase with ideal three-flavor symmetry, is applied to
two-flavor matter.
Our observation is that neutron star matter can continuously be
connected to two-flavor color-superconducting (2SC) phase with an
additional condensate of $d$-quarks in the $^3P_2$ channel.
I discuss here two aspects of this novel phase.
First, I introduce the notion of continuity based on the patterns of
symmetry breaking and the corresponding order parameters, and then
explain qualitatively the physics mechanism of $d$-quark $^3P_2$
pairing in analogy to nuclear physics.
Our finding serves as the theoretical underpinning for the
phenomenological construction of the equation of state in the neutron
star environment.
\end{abstract}


\end{frontmatter}


\section{Introduction}

Whether there is a phase boundary between hadronic matter and quark
matter at high densities has been a long-standing problem since the
advent of QCD.
I shall present here a possibility of continuous connection between
hadronic matter and color-superconducting quark matter, based on our
recent work~\cite{Fujimoto:2019sxg}.
The idea of quark-hadron continuity is ascribed to the identical
symmetry breaking patterns and the low-lying excitations in both quark
and hadronic phase.
In three-flavor symmetric case, where color-flavor locking (CFL)
occurs, continuity is established~\cite{Schafer:1998ef}.
By contrast, in two-flavor case, the concept of continuity
between nuclear matter and two-flavor color superconductor (2SC) is
generally thought to be invalidated due to the apparent difference in
the global symmetries realized in the both phases.
In this contribution, I shall concentrate on the two-flavor case and
discuss that the continuity from $^3 P_2$ neutron superfluid to
two-flavor quark matter is still possible with an additional down
quark condensate in $^3 P_2$ channel~\cite{Fujimoto:2019sxg}.

It is worth mentioning that quark-hadron continuity is significant in
cold and dense QCD matter, especially when we consider the phase
structure and the bulk properties.
A Ginzburg-Landau analysis shows that matter at sufficiently low
temperature allows for a critical point in the phase diagram, and goes
through a smooth crossover from the hadronic to the quark phase with
increasing density (see, e.g., Ref.~\cite{Hatsuda:2006ps}).
The bulk properties of dense matter is characterized by a quantity
such as the equation of state (EoS).
Now quark-hadron continuity is the baseline for the phenomenological
construction of EoS~\cite{Baym:2017whm}.

\section{Quark-hadron continuity --- aspects of symmetry and order parameter}

In the conventional color superconductor, a diquark condensate in $^1
S_0$ channel is formed
\begin{equation}
  \langle \hat{q}^\top_{\alpha A} \mathcal{C} \gamma^5 \hat{q}_{\beta B} \rangle
  \propto \epsilon_{\alpha \beta \gamma} \epsilon_{ABC}
  \langle\hat{\Phi}^{\gamma C} \rangle\,,
  \label{eq:qq}
\end{equation}
where the charge conjugation matrix $\mathcal{C} \equiv i \gamma^0
\gamma^2$ is inserted to form a Lorentz scalar.
Greek $(\alpha,\beta,\ldots)$ and capital $(A,B,\ldots)$ indices
represent color and flavor, respectively.
The color-superconducting phase can be thought of as a Higgs phase of
QCD.
Diquarks play here the same role as the Higgs boson does in the
electro-weak sector of the standard model.
It has previously been shown in lattice gauge theories that
the confinement phase and the Higgs phase are connected without a
phase boundary~\cite{Fradkin:1978dv}.
Owing to this fact, we cannot distinguish between the hadronic
(confinement) phase and the color superconductor (Higgs) phase as long
as the global symmetries are the same in both phases;
it leads to the concept of \textit{quark-hadron continuity}.

The symmetry of QCD comprises local gauge symmetry, chiral symmetry
and baryon number symmetry: $\mathcal{G}_{\rm QCD} =
[\mathrm{SU}(3)_{\rm C}] \times \mathrm{SU}(N_f)_{\rm L} \times
\mathrm{SU}(N_f)_{\rm R} \times \mathrm{U(1)}_{\rm B}$.
Neglecting the strange ($s$) quark mass, the chiral part of
$\mathcal{G}_{\rm QCD}$ can be ideally treated as $N_f = 3$.
In reality, however, $s$ is much heavier than the up ($u$) and the
down ($d$) quarks ($m_s / m_{u,d} \sim 30$), so it is more natural to
consider isospin-symmetric $N_f =2$ systems.

Here, I first review the conventional continuity in $N_f = 3$ from the
viewpoint of residual symmetry in the color-superconducting phase and
the corresponding order parameters.
Then I will show our novel analysis in $N_f=2$ case, which also allows
for continuity with a new element of $^3 P_2$ $d$-quark condensate.

\subsection{$N_f = 3$ case}

In this case, color-flavor locking occurs.
The CFL ansatz for the diquark condensate Eq.~\eqref{eq:qq} reads
$\langle \hat{\Phi}^{\alpha A}\rangle = \delta^{\alpha A} \Delta_{\rm
  CFL}$.
The residual symmetry in the CFL phase is $\mathcal{G}_{\rm CFL} =
\mathrm{SU(3)_{C+L+R}}$.
Thus, the pattern of symmetry breaking in the CFL phase is
$\mathcal{G}_{\rm QCD} \to \mathcal{G}_{\rm CFL}$.
The residual global symmetry in the CFL phase $\mathcal{G}_{\rm CFL}$
is the same as that of the hadronic phase, where the chiral and the
baryon number symmetries are spontaneously broken.
Thus, there exists quark-hadron continuity for $N_f = 3$ case.

As previously proven in Ref.~\cite{Elitzur:1975im}, local gauge
symmetry cannot be broken spontaneously.
It means that the Higgs mechanism alone, which is \textit{fictitious}
breaking of gauge symmetry, cannot be captured by any gauge-invariant
order parameter.
Here, however, global symmetry breaking occurs simultaneously in
$\mathcal{G}_{\rm QCD} \to \mathcal{G}_{\rm CFL}$, thus this global
sector can still be captured by gauge-invariant order parameters.
We can construct such order parameters by saturating gauge indices
of diquark operator~\eqref{eq:qq} in two ways, each of which captures
the breaking of the chiral and superfluid sector of the QCD symmetry
$\mathcal{G}_{\rm QCD}$~\cite{Rajagopal:2000wf}.
\begin{align}
  \text{Chiral}: \hspace{10pt} \mathcal{M} &=
  \delta^{\beta}_{\alpha} \delta^{\beta'}_{\alpha'} (\bar{q}^{\alpha}
  \bar{q}^{\alpha'}) (q_{\beta} q_{\beta'}) 
  \propto (\bar{q}^{\alpha} q_{\alpha})(\bar{q}^{\alpha'}q_{\alpha'})\,,\\
  \text{Superfluid}: \hspace{14pt} \Upsilon &= \epsilon^{\alpha \beta \gamma} \epsilon^{\alpha' \beta'
  \gamma'} (q_{\alpha} q_{\alpha'}) (q_{\beta} q_{\beta'}) (q_{\gamma}
  q_{\gamma'})
  \propto (\epsilon^{\alpha\beta\gamma} q_\alpha q_\beta q_\gamma)
  (\epsilon^{\alpha'\beta'\gamma'} q_{\alpha'} q_{\beta'}
  q_{\gamma'})\,.
  \label{eq:lambda}
\end{align}
Superfluid order parameter $\Upsilon$ takes the similar expectation
value in both the quark and the hadronic phases because of
quark-hadron continuity.
In the hadronic phase, as it is obvious from the RHS of
Eq.~\eqref{eq:lambda} above, it can be interpreted as $\Upsilon
\propto \Lambda\Lambda$.
This is the order parameter for $\Lambda$-superfluidity.

\subsection{$N_f = 2$ case}

The 2SC ansatz for the diquark condensate~\eqref{eq:qq} reads
\begin{equation}
\langle \hat{\Phi}^{\alpha A}\rangle = \delta^{\alpha 3} \delta^{A 3}
\Delta_{\rm 2SC} \equiv  \Phi^{\alpha}_{\rm 2SC}\,,
\label{eq:2SC}
\end{equation}
where we assumed the unitary gauge-fixing.
The residual symmetry in the 2SC phase is $\mathcal{G}_{\rm 2SC} =
[\mathrm{SU(2)_C}] \times \mathrm{SU(2)_L} \times \mathrm{SU(2)_R}
\times \mathrm{U(1)_B}$, which is apparently different from the
hadronic phase.
However, apart from the 2SC condensate $\Phi^\alpha_{\rm 2SC}$, it is
still possible to consider an additional two-flavor diquark condensate
in order that quark-hadron continuity is maintained.
To find a new element, here I turn the three-flavor argument the other
way round, focusing on the superfluid aspect.
We describe neutrons by a quark-diquark structure $n =
\epsilon^{\alpha\beta\gamma} (u_\alpha^\top \mathcal{C} \gamma^5 d_\beta)
d_\gamma$.
Assuming quark-hadron continuity, which will be justified \textit{a
  posteriori}, we start out with the order parameter for $^3 P_2$
neutron superfluidity:
\begin{equation}
  \Upsilon_{nn} \equiv n^\top \mathcal{C} \gamma^i \nabla^j n \propto
  \epsilon^{\alpha \beta \gamma} \epsilon^{\alpha' \beta' \gamma'}
  (u^\top_\alpha \mathcal{C} \gamma^5 d_{\beta})
  (u^\top_{\alpha'} \mathcal{C} \gamma^5 d_{\beta'})
  (d^\top_\gamma \mathcal{C} \gamma^i \nabla^j d_{\gamma'})\,.
\end{equation}
The first two terms in parentheses are identical to the 2SC condensate
$\Phi^\gamma_{\rm 2SC}$ in Eq.~\eqref{eq:2SC}, while the last term
$d^\top_{\gamma} \mathcal{C} \gamma^i \nabla^j d_{\gamma'}$ is the
novel element here.
If we take the expectation value of $\Upsilon_{nn}$ in the mean-field
approximation, $\Upsilon_{nn}$ takes the value
\begin{equation}
  \langle \Upsilon_{nn} \rangle \approx \Phi^{\alpha}_{\rm 2SC}
  \Phi^{\alpha'}_{\rm 2SC} \langle d^\top_{\alpha} \mathcal{C} \gamma^i
  \nabla^j d_{\alpha'}\rangle\,,
\end{equation}
where we have neglected the crystalline condensation.
We call this phase 2SC+$\dd$ phase.
The residual symmetry in the 2SC+$\dd$ phase is the same as neutron
matter, hence the continuity holds.
At first glance, $d^\top_{\alpha} \mathcal{C} \gamma^i \nabla^j
d_{\alpha'}$ condensation seems not to occur because $d$-quarks are in
color $\boldsymbol{6}$ channel, which is known to be repulsive in one gluon
exchange (OGE) potential.
Even though the short-range interaction via OGE is repulsive, $^3P_2$
pairing is still possible through the spin-orbit interaction as I will
explain shortly.

\section{Justifying the 2SC+$\dd$ phase}
\subsection{Dynamical mechanism favoring $d$-quark pairing in $^3P_2$ channel}

A prototype example of the $^3 P_2$ superfluid is realized for
neutrons inside neutron stars.
Neutrons undergo pairing in a $^1 S_0$ state at $n_{\rm B} < 0.5\,n_0$ (with
$n_0 \simeq 0.16 \,\text{fm}^{-3}$, the normal nuclear density) and in the
$^3P_2$ state at $n_{\rm B} > n_0$.
One can find $^1S_0$ and $^3P_2$ neutron superfluid in the inner crust
and the outer core of neutron stars, respectively.
This realization of $^3P_2$ superfluidity is based on the observed
pattern of nucleon-nucleon ($NN$) scattering phase
shifts~\cite{Hoffberg:1970vqj,Tamagaki:1970ptp}.
The phase shift of the $^1S_0$ partial wave changes sign from positive
to negative with increasing energy of the two nucleons, indicating
that the pairing interaction turns from attractive to repulsive with
increasing Fermi energy.
Consequently, pairing in the $^1S_0$ channel is disfavored at high
densities and taken over by pairing in the $^3P_2$ channel.
This property is attributed to the short-repulsive core in the singlet
$S$-wave and the significant attraction selectively generated by the
spin-orbit interaction in the triplet $P$-wave with total angular
momentum $J = 2$.
The selective attraction for $J=2$ channel is generated through the factor
$-\boldsymbol{L}\cdot\boldsymbol{S} = -[J(J+1) - L(L+1) - S(S+1)]/2$
in the spin-orbit potential, which are $+2$, $+1$ and $-1$ in $^3P_0$,
$^3P_1$ and $^3P_2$ states, respectively.
With an extra minus sign in the spin-orbit potential, there is
attraction only in $^3P_2$ channel.
All other $S$- and $P$-wave $NN$ phase shifts in isospin $I=1$ channel
are smaller or repulsive in matter dominated by neutrons.

The same mechanism applies in quark matter favoring $^3P_2$ pairing
between $d$-quarks as specifically described in
Ref.~\cite{Fujimoto:2019sxg}.
Short-range repulsion arises between $d$-quarks due to the repulsive
nature of OGE in color $\boldsymbol{6}$ channel.
Also, near the confinement region in the continuity picture,
short-distance repulsive interaction between quarks can be thought of
as emerging from quark-gluon exchange in a non-relativistic quark
model picture.
Spin-orbit force in a long range also arises between quarks.
One can derive it employing Nambu--Jona-Lasinio (NJL) model with
vector coupling with a sizable effect.
This effect could also be understood as naturally emerging from the
Fermi-Breit reduction of the OGE potential.
Therefore, pairing between $d$-quarks in the $^1S_0$ channel is also
disfavored and taken over by pairing in the $^3P_2$ channel,
justifying our picture with 2SC+$\dd$ phase.

\subsection{Relation with the dense matter EoS and neutron star phenomenology}

In the context of previous mean-field calculations of
color-superconducting quark matter in an NJL-type model, a
four-fermion coupling term in the $^3P_2$ channel has so far been
missing.
It would then be instructive to see how the interaction in
this channel could be enlarged through the coupling to the
energy-momentum tensor.
Considering the four-fermion coupling term in the $^3 P_2$ diquark channel,
i.e.,
\begin{align}
  \hat{\cal I}_P &= (\bar{\psi} \gamma^{i} \nabla^{j} \mathcal{C} \bar{\psi}^\top)
            (\psi^\top \mathcal{C} \gamma_{i} \nabla_{j} \psi)\,.
  \label{eq:IP-diquark}
\end{align}
Fierz rearranging this $\hat{\cal I}_P$ term, one finds a term with a
direct correspondence to the energy-momentum tensor in the fermionic
sector, $T^{\mu\nu}=\bar{\psi}i\gamma^\mu\partial^\nu\psi$.
For matter in equilibrium, $T^{\mu\nu}=\mathrm{diag}[\varepsilon,P,P,P]$,
with the energy density $\varepsilon$ and the pressure $P$ of
fermionic matter.
Extracting only the contribution relevant for the energy-momentum tensor,
the tree-level expectation value of $\hat{\cal I}_P$ reads
\begin{equation}
  \langle\hat{\cal I}_P\rangle \approx \frac{3}{4}P^2 \,.
  \label{eq:vevint}
\end{equation}
It is evident from this expression that the $^3 P_2$ diquark
interaction couples to the pressure which is a macroscopic quantity.
Even if the direct mixing between the quark-antiquark (hole) and the
diquark sectors may not be large, the superfluid energy gap can be
enhanced by the macroscopic expectation value of the energy-momentum
tensor as given in Eq.~\eqref{eq:vevint}.

If we regard $P$ here as a function of the energy density
$\varepsilon$, it is nothing but the EoS $P=P(\varepsilon)$.
Quark-hadron continuity scenario is now also relevant in the context
of neutron stars, particularly for the phenomenological construction
of the dense matter EoS~\cite{Baym:2017whm} as is evident from the
direct correspondence in the expression of Eq.~\eqref{eq:vevint}.
The EoS is constructed within three-window modeling, where the
boundary conditions are set by the nuclear EoS at $n_{\rm B} <
2\,n_0$ and the quark EoS at $n_{\rm B} > 5\,n_0$.
The intermediate region is interpolated with smooth function, whose
physics background is quark-hadron continuity (see, e.g., Fig.~4 of
Ref.~\cite{Baym:2017whm} for an illustration of the three-window
modeling).
Our two-flavor scenario here gives more convincing baseline for the
three-window modeling than the three-flavor scenario, as most of the
nuclear EoS at $n_{\rm B} \le 2n_0$ is calculated for nucleon degrees
of freedom.
Conversely, recent attempts to extract the neutron star EoS directly
from astrophysical observations, such as that utilizing deep neural
network~\cite{Fujimoto:2017cdo,Fujimoto:2019hxv}, may provide a basis
for judging the continuity hypothesis in the future.

\section{Summary}

In this work, we showed that neutron matter realized inside neutron
stars are continuously connected to two-flavor color superconductor.
We found that the quark phase counterpart of the neutron $^3P_2$
superfluid is what we call 2SC+$\dd$ phase, which is a 2SC phase with
$\dd$ pairing in $^3 P_2$ channel.
The mechanism favoring this phase is two-fold:
One is that the short-range repulsive force emerging from quark-gluon
exchange, which can also be interpreted with OGE potential.
The other is that the long-range attraction arising from the
spin-orbit configuration favoring $J=2$ in particular.

\bigskip
\bigskip

\textbf{Acknowledgments.}---
I thank Kenji~Fukushima and Wolfram~Weise for fruitful collaborations
and encouraging me to write this contribution on my own account.
Also, the author is grateful to Toru~Kojo for discussions.





\bibliographystyle{elsarticle-num}
\bibliography{continuity}

\begin{thebibliography}{10}
\expandafter\ifx\csname url\endcsname\relax
  \def\url#1{\texttt{#1}}\fi
\expandafter\ifx\csname urlprefix\endcsname\relax\def\urlprefix{URL }\fi
\expandafter\ifx\csname href\endcsname\relax
  \def\href#1#2{#2} \def\path#1{#1}\fi

\bibitem{Fujimoto:2019sxg}
Y.~Fujimoto, K.~Fukushima, W.~Weise, {Continuity from neutron matter to
  two-flavor quark matter with $^1 S_0$ and $^3 P_2$ superfluidity}, \href
  {http://arxiv.org/abs/1908.09360} {\path{arXiv:1908.09360}}.

\bibitem{Schafer:1998ef}
T.~Sch{\" a}fer, F.~Wilczek, {Continuity of quark and hadron matter}, Phys.
  Rev. Lett. 82 (1999) 3956--3959.
\newblock \href {http://arxiv.org/abs/hep-ph/9811473}
  {\path{arXiv:hep-ph/9811473}}.

\bibitem{Hatsuda:2006ps}
T.~Hatsuda, M.~Tachibana, N.~Yamamoto, G.~Baym, {New critical point induced by
  the axial anomaly in dense QCD}, Phys. Rev. Lett. 97 (2006) 122001.
\newblock \href {http://arxiv.org/abs/hep-ph/0605018}
  {\path{arXiv:hep-ph/0605018}}.

\bibitem{Baym:2017whm}
G.~Baym, T.~Hatsuda, T.~Kojo, P.~D. Powell, Y.~Song, T.~Takatsuka, {From
  hadrons to quarks in neutron stars: a review}, Rept. Prog. Phys. 81~(5)
  (2018) 056902.
\newblock \href {http://arxiv.org/abs/1707.04966} {\path{arXiv:1707.04966}}.

\bibitem{Fradkin:1978dv}
E.~H. Fradkin, S.~H. Shenker, {Phase Diagrams of Lattice Gauge Theories with
  Higgs Fields}, Phys. Rev. D19 (1979) 3682--3697.

\bibitem{Elitzur:1975im}
S.~Elitzur, {Impossibility of Spontaneously Breaking Local Symmetries}, Phys.
  Rev. D12 (1975) 3978--3982.

\bibitem{Rajagopal:2000wf}
K.~Rajagopal, F.~Wilczek, {The Condensed matter physics of QCD}, in:
  M.~Shifman, B.~Ioffe (Eds.), At the frontier of particle physics. Handbook of
  QCD. Vol. 1-3, 2000, pp. 2061--2151.
\newblock \href {http://arxiv.org/abs/hep-ph/0011333}
  {\path{arXiv:hep-ph/0011333}}.

\bibitem{Hoffberg:1970vqj}
M.~Hoffberg, A.~E. Glassgold, R.~W. Richardson, M.~Ruderman, {Anisotropic
  Superfluidity in Neutron Star Matter}, Phys. Rev. Lett. 24~(14) (1970) 775.
\newblock \href {http://dx.doi.org/10.1103/PhysRevLett.24.775}
  {\path{doi:10.1103/PhysRevLett.24.775}}.

\bibitem{Tamagaki:1970ptp}
R.~Tamagaki, \href{http://dx.doi.org/10.1143/PTP.44.905}{{Superfluid State in
  Neutron Star Matter. I. Generalized Bogoliubov Transformation and Existence
  of ${}^3P_2$ Gap at High Density}}, Prog. Theor. Phys. 44~(4) (1970)
  905--928.

\bibitem{Fujimoto:2017cdo}
Y.~Fujimoto, K.~Fukushima, K.~Murase, {Methodology study of machine learning
  for the neutron star equation of state}, Phys. Rev. D98~(2) (2018) 023019.
\newblock \href {http://arxiv.org/abs/1711.06748} {\path{arXiv:1711.06748}}.

\bibitem{Fujimoto:2019hxv}
Y.~Fujimoto, K.~Fukushima, K.~Murase, {Mapping neutron star data to the
  equation of state of the densest matter using the deep neural network}, \href
  {http://arxiv.org/abs/1903.03400} {\path{arXiv:1903.03400}}.

\end{thebibliography}







\end{document}